# A Study of a Mini-drift GEM Tracking Detector

B.Azmoun, B.DiRuzza, A.Franz, A.Kiselev, R.Pak, M.Phipps, M.L.Purschke, C.Woody

*Abstract*— A GEM tracking detector with an extended drift region has been studied as part of an effort to develop new tracking detectors for future experiments at RHIC and for the Electron Ion Collider that is being planned for BNL or JLAB. The detector consists of a triple GEM stack with a small drift region that was operated in a mini TPC type configuration. Both the position and arrival time of the charge deposited in the drift region were measured on the readout plane which allowed the reconstruction of a short vector for the track traversing the chamber. The resulting position and angle information from the vector could then be used to improve the position resolution of the detector for larger angle tracks, which deteriorates rapidly with increasing angle for conventional GEM tracking detectors using only charge centroid information. Two types of readout planes were studied. One was a COMPASS style readout plane with 400 μm pitch XY strips and the other consisted of 2x10mm$^2$ chevron pads. The detector was studied in test beams at Fermilab and CERN, along with additional measurements in the lab, in order to determine its position and angular resolution for incident track angles up to 45 degrees. Several algorithms were studied for reconstructing the vector using the position and timing information in order to optimize the position and angular resolution of the detector for the different readout planes. Applications for large angle tracking detectors at RHIC and EIC are also discussed.

## I. INTRODUCTION

GEM detectors are widely used in many tracking applications in high energy and nuclear physics. They typically provide two dimensional coordinate information using a segmented strip or pad readout plane, or can be used in a TPC configuration where the drift time of the collected charge can be used to determine the third position coordinate. We have studied a hybrid of these two configurations which we call a minidrift GEM detector, where we have introduced a moderate size drift region above the GEM stack to collect the charge deposited by particles traversing this region. By measuring the drift time of the ionization clusters, one can determine the angle of the track passing through the detector. The position and angle of the track can be used to define a vector which provides a substantial improvement in position resolution at larger incident angles compared to a simple centroid measurement. This allows a reduction in the number of measuring stations required to measure tracks to a given precision, which in turn reduces the amount of material in the particle's path. All of these requirements are important for future tracking detectors at RHIC, and in particular, at a future Electron Ion Collider (EIC), where achieving high resolution with a minimal amount of material for the scattered electron is important.

We have investigated two readout structures for the detector which could be used for different applications depending on the particle multiplicty. The first is a COMPASS style readout with 400 μm pitch strips in the X and Y directions [1]. For low multiplicities, this type of readout has been used to provide excellent position resolution for small angle tracks at very high rates. It may therefore also be suitable for EIC where particle multiplicities in the direction of the scattered electron are also low. However, for high multiplicity events, such as in heavy ion collisions, a two coordinate XY readout cannot be used due to the large number of ambiguities produced by multiple tracks in the same region of the detector. In this situation, two dimensional pad readouts are typically used, but to achieve good spatial resolution, a large number of small pads are required. Alternatively, a chevron style readout [2,3] can be used with relatively large pads (~ few mm) which exploit the charge sharing between interspersed electrodes within the chevron to achieve a resolution that is much smaller than the pad size. This type of readout has also been used for TPCs where a high degree of pad segmentation is required [4,5]. We have studied the minidrift GEM detector with a 2x10 mm$^2$ chevron pad readout, where fine chevon strips along the 2 mm direction provided precise position information, and the 10 mm dimension was chosen simply for segmentation purposes.

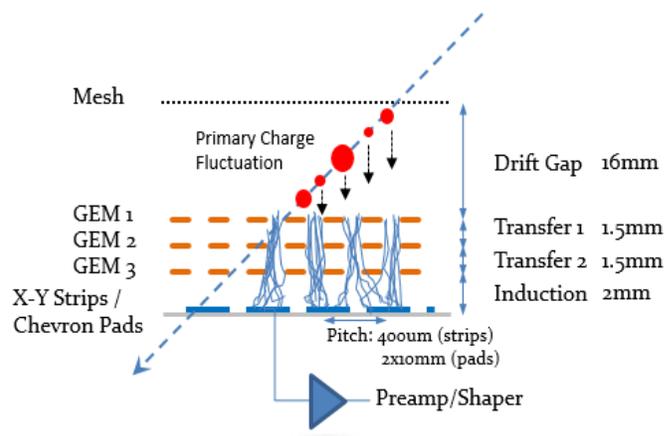

Fig. 1. Illustration of the minidrift detector consisting of a triple GEM stack with a 16 mm drift gap above and either XY strips or chevron pad readout.

Manuscript submitted on August 24, 2015. This work was supported in part by the U.S. Department of Energy under Prime Contract No. DE-SC0012704.

B.Azmoun, B.DiRizza, A.Franz, A.Kiselev, R.Pak, M.Phipps, M.L.Purschke, and C.Woody are with Brookhaven National Laboratory, Upton, NY.

Figure 1 shows a conceptual illustration of how the minidrift detector works. The incident track deposits charge along its path in the drift gap which then drifts to a three stage GEM detector below. The initial charge deposit is subject to fluctuations in the primary cluster formation, which can be quite large and can ultimately limit the position resolution. The charge undergoes diffusion in the drift gap and in the amplification process in the GEM detector and is finally deposited on the readout electrodes, which are either strips or chevron pads.

Our studies of the minidrift detector have included extensive measurements in the lab as well as several beam tests, and the results from several of these studies have already been published [6,7]. In this paper, we concentrate on the results from the latest beam tests at Fermilab where we measured the position and angular resolution of both the XY strip and chevron readouts over a range of angles from normal incidence up to 45°. These measurements represent an improvement over our previous results measured at CERN in that we utilized a silicon telescope to determine the incoming beam track with very high precision which then allowed us to measure the position and angular resolution of the minidrift detector with much greater accuracy. We will describe the experimental setup used to study the detectors, the analysis procedure used to determine their spatial and angular resolution, and possible applications of these types of detectors at RHIC and EIC.

## II. EXPERIMENTAL SETUP

The minidrift detector consisted of a stack of three standard 10 x 10 $cm^2$ CERN GEM foils with a 16 mm drift gap above. Two separate detectors were constructed, one with a COMPASS style readout plane and the other with 2 x 10 $mm^2$ chevron pads. The structure of the COMPASS readout plane is shown in Fig. 2a, which has two independent sets of strips separated by a kapton layer, each with a pitch of 400 μm. A detailed description of this readout structure is given in Ref. [1]. Figure 2b shows the chevron structure which has interleaving zigzag electrodes with a pitch of 0.5 mm and a pad spacing of 2 mm. The length of the pads was 10 mm and was chosen simply as a convenient segmentation. Charge deposited on the chevrons is typically shared unequally between the electrodes and the position within the pad can be determined by computing a weighted centroid. A more detailed description of how the chevron readout functions is given in Refs. [2,3].

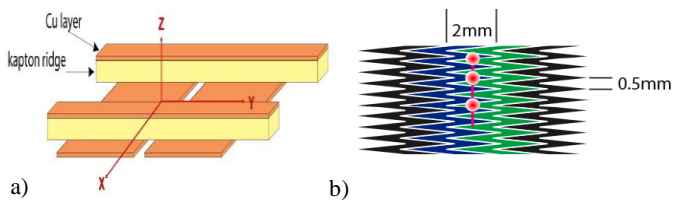

Fig. 2 a) COMPASS readout with 80 μm wide top strips and 350 μm wide bottom strips separated by a 50 μm kapton layer, each with a pitch of 400 μm (see Ref [1]) b) Chevron readout structure with interleaving zigzag electrodes with a pitch of 0.5 mm and 2 mm spacing.

The detectors were operated with a mixture of $Ar/CO_2$ (70/30) at atmospheric pressure at a gain ~ few x $10^3$. In the case of the chevron readout, which collects significantly more charge per pad than the strips, some saturation of the preamp signals occurred for large pulses. However, as discussed in the next section, this did not have a significant effect on the measured resolution. The readout used the CERN SRS system [8] which digitized the signals from each pad or strip in 25 ns time bins. The SRS system was read out using our own RCDAQ data acquisition system [9].

The detectors were tested in the MT6 test beam at Fermilab in October 2013 and February 2014. The MT6 test beam area was instrumented with a high precision silicon telescope which provided an independent measurement of the incoming beam track on an event by event basis. The GEM detector was placed ~ 35 cm from the downstream end of the telescope on a rotatable stand that allowed changing the angle of the detector with respect to the incoming beam around its vertical axis. The test beam setup is shown in Fig.3. The silicon telescope was read out using its own data acquisition system which provided an independent set of events that had to be synchronized with the events from the GEM readout in the offline analysis.

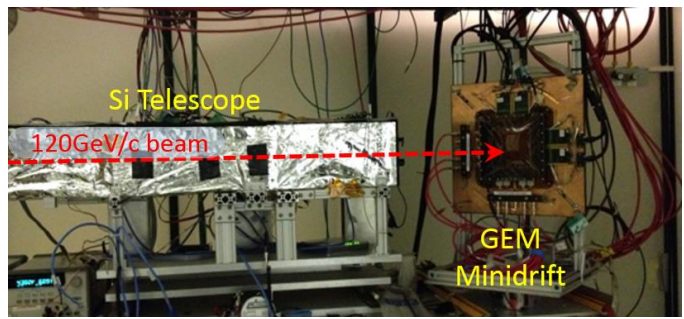

Fig. 3 Setup in the MT6 test beam area at Fermilab showing the silicon telescope and the minidrift GEM detector on its rotatable stand.

## III. DATA ANALYSIS

The analysis of the test beam data utilized the silicon telescope to measure the incoming beam track and compute a projected position at the readout plane of the GEM detector. The silicon telescope was part of the MT6 beam line instrumentation and consisted of four 1 $cm^2$ silicon pixel detectors that provided the parameters of the beam track with a spatial resolution ~ 17 μm and an angular resolution ~ 10 μrad. The data from the telescope was processed run by run using its own analysis software package and the events were correlated with the events taken with the GEM data acquisition system in the offline analysis.

The data from the GEM detector consisted of pulse height information from each readout channel sampled every 25 ns using the APV25 readout chip with the SRS DAQ system. This provided the charge and time information used to reconstruct a vector that was used to determine the position and angle of the track. Figure 4a shows an example of the raw digitized data from several strips for a track incident at an inclined angle and shows the long shaping time (~ 75 ns) of the readout electronics.

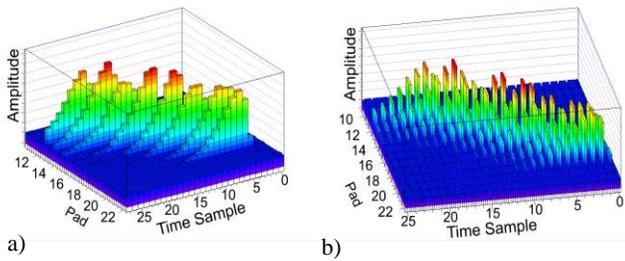

Fig. 4 a) Event display of digitized preamp signals from several strips for an inclined track passing through the detector. b) Example of the charge distributions used to compute a centroid in each time bin for a group of strips for an inclined track.

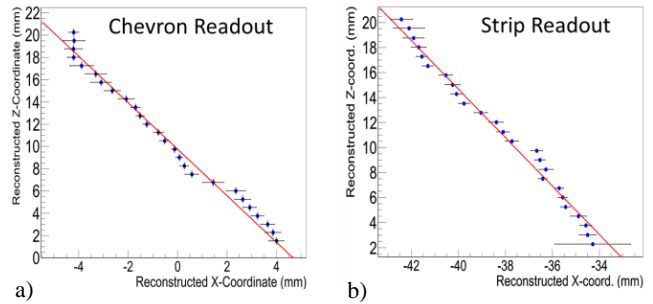

Fig. 5. Examples of a vector reconstructed in the X-Z using the time slice method: a) strip readout b) chevron pad readout.

The position and angle of the track were reconstructed using a *time slice method* which essentially treats the data in each time sample independently. We also studied a rising edge method which fits the rising edge of the pulse to determine the arrival time of the charge on each electrode. This method is described in Refs. [6,7] and a comparison of the results obtained with these two methods is in Section 4.3. With the time slice method, a charge weighted centroid was calculated for each time bin using all channels with a pulse height above a given threshold. The centroid coordinates were then used along with the average values of the time bins to compute a vector. An example of the charge distributions used to compute the centroids for the strips is shown in Fig. 4b. In the case of the strips, this was done independently in the X-Z and Y-Z planes. For the chevron pads, a vector was only computed in the X-Z plane, which gave the position coordinate in the chevron direction (i.e., along the 2 mm direction of the interspersing zigzag strips). Because of the much coarser (10 mm) spacing of the pads in the Y direction, no vector was computed in the Y-Z plane for the chevrons. An example of a vector reconstructed in the X-Z plane for both the strips and chevron pads is shown in Fig. 5.

The gain of the detector was kept at a few x $10^3$ when measuring the chevron pads. This helped improve the centroid measurement by keeping smaller signals above the noise, but due to the limited dynamic range of the preamp in the APV25 readout chip, it led to some saturation for larger signals near the center of the centroid. However, since the tails of the charge distribution have a much larger influence on the determination of the position from the centroid measurement, and the signals in the tails did not suffer from saturation, the saturation of the signals near the center did not significantly affect the resolution. On the other hand, due to the long decay time of the preamp pulse, the tails from neighboring pads did affect the centroid determination in each time bin using the time slice method. This was difficult to correct for using a simple analysis, but we have studied a more sophisticated track reconstruction algorithm that unfolds the effect of the pulse shape of the preamp which we believe is capable of achieving much better position resolution. This method will be discussed in Section 4.4.

For the strip readout, the detector was measured in three different orientations of the readout plane. Most of the data was taken in what we called the "diamond position" where the detector was rotated by 45° in a plane perpendicular to the incoming track, as denoted by the ϕ angle shown in Fig. 6. This gave roughly equal contributions to the resolution from the X and Y coordinates, although the X strips (which are the lower strips in the COMPASS readout) collected ~ 30% less charge and gave somewhat poorer resolution. We take the overall position resolution as the average quadrature sum of the resolutions of the X and Y coordinates measured separately: $\sigma_{ave} = \sqrt{(\sigma_x^2 + \sigma_y^2)/2}$ . The detector was then rotated about its vertical axis to study the dependence of the resolution as a function of this angle, which we define as θ. We also studied the detector with both the X and Y strips oriented along the vertical axis (which we call the orthogonal position) that gave an independent measure of the resolution for the two coordinates separately. For the pad readout, the detector was only measured in one orientation, which was with the chevrons aligned in the horizontal direction (i.e., perpendicular to the axis of rotation), which gave a measure of the chevron resolution as a function of the rotation angle.

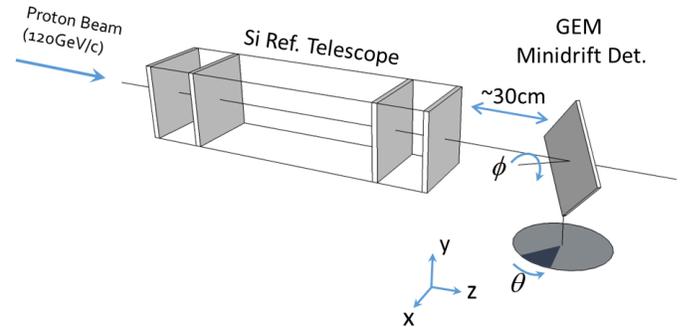

Fig. 6 Orientation of the GEM detector with respect to Si telescope. The angle θ is the angle of incidence of the incoming beam.

The resolution of the GEM detector was determined by comparing the position and angle of the track from the GEM with the projected track from the silicon telescope at a reference plane inside the GEM detector. Due to the much higher precision of the silicon telescope, its resolution was negligible (~ 17μm) at the reference plane of the GEM compared with that of the GEM itself. However, the coordinate system for the silicon telescope and the internal coordinate system of the GEM had to be carefully aligned with respect to each other, both in position and angle, in order to determine the GEM detector resolution.

The GEM was aligned with the silicon telescope by minimizing the residual distributions of the projected track from the telescope and the track found by the GEM at a common reference plane, which was taken to be the GEM readout plane. Figure 7 shows some typical residual distributions for both position and angle for the strip readout. At each incident angle, the residual distributions were fit to a double Gaussian over their entire range in order to determine the background contribution to the width. The detector resolution was taken to be the sigma of the narrower Gaussian and the broader distribution was taken as background. In the case of the angular resolution, a slight asymmetry in the residual distribution was observed. There were no quality cuts applied to the tracks at this stage of the analysis, and the detector efficiency was generally greater that 90% at all the angles. Further details of this alignment procedure are discussed in [6].

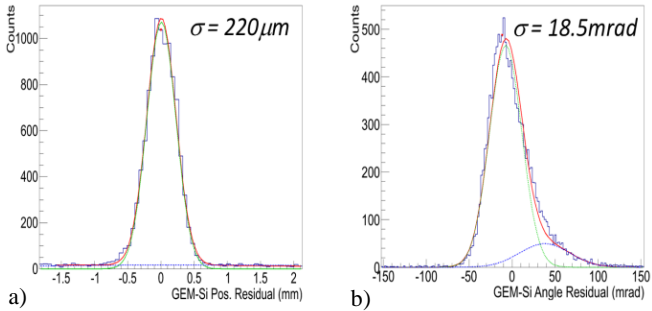

Fig. 7. Typical residual distributions for position and angle for the strip readout along with a double Gaussian fit (red). The detector resolution is derived from the narrower Gaussian (green), while the broader Gaussian (blue) is treated as background.

One inherent feature of the SRS readout system was that the phase of the 40 MHz clock was not synchronized with the trigger input. This lead to a timing uncertainty of ~ 25 ns/√12 (~ 7 ns) in the measured arrival time of the charge on each electrode. Since this time shift was the same for all channels, it did not affect the angle of the reconstructed vector, but it did result in a random smearing of the intercept position of the vector on the reference plane. This could not be corrected for on an event by event basis and had to be unfolded from the measured position resolution. However, this timing uncertainty did not affect the measured angular resolution.

## IV. RESULTS

### 4.1 COMPASS Strips

Figure 8 shows the position resolution of the minidrift detector with the COMPASS readout strips measured in the three different orientations as a function of the angle θ with respect to the incoming beam. In the two orthogonal positions, the angle goes out to 45°, which was the angle the detector made with the beam, whereas for the diamond position, the projected angle in each plane only goes out to ~ 32°.

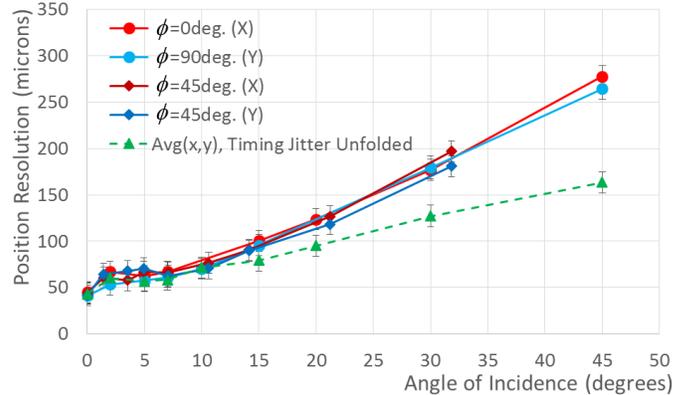

Fig. 8. Position resolution for the COMPASS strips measured in three different orientations as a function of angle with respect to the incoming beam. Also shown is the resolution after unfolding the contribution from the timing jitter.

The measured resolution at each detector orientation agrees very well. It starts at a value ~ 50 μm at zero degrees and reaches a value ~ 275 μm at 45°. However, after unfolding the contribution from the timing jitter, the resolution decreases to ~ 160 μm at 45°.

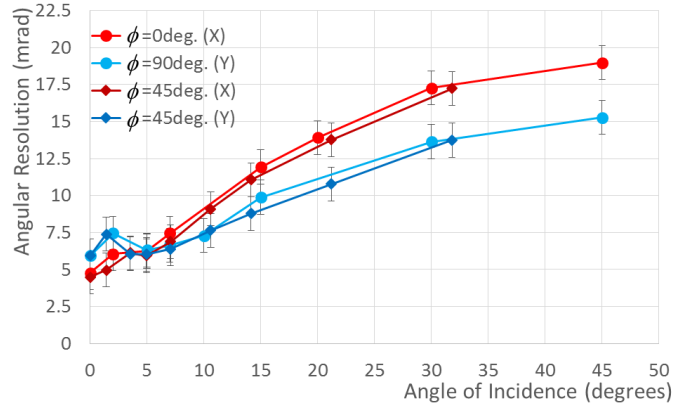

Fig. 9. Angular resolution for the COMPASS strips measured in the three different orientations as a function of angle with respect to the incoming beam

Figure 9 shows the angular resolution of the vector reconstructed with the strip readout as a function of angle. The angular resolution is ~ 5-6 mrad at small angles and increases to ~ 15-20 mrad at 45°. The resolution for the X strips is consistently worse than for the Y strips, both in the orthogonal position and in the diamond position. We believe this is due to the fact that the X strips collect ~ 30% less charge than the Y strips.

*4.2 Chevron Pads*

The chevron pads only provided a high resolution coordinate measurement along the direction of the interspersed zigzag strips (i.e., the horizontal direction shown in Fig. 2). Therefore, the resolution quoted is the resolution measured along this direction. In the other direction, the resolution was essentially just given by the pad size (2 mm/√12 ~ 600 μm). This would be similar to the way such a readout plane would be used in a detector where one is measuring a high precision coordinate in the bending direction inside a magnetic field, and the segmentation for the other coordinate is chosen to reduce occupancy or channel count.

The chevron pads produce an inherently non-uniform response at the level of the pad spacing due to the fact that the charge spread is measured with finite resolution with the zigzag strips. This was studied in the lab using a high precision X-ray source to measure the uniformity of response across the zigzag direction. The X-ray source was equipped with a collimator with a 50 μm slit that produced a narrow beam of X-rays that was scanned across the pad plane in 100 μm steps using an automated stepping motor. The position computed from the centroid measurement from the pads was compared with the known position of the source from the stepping motor and used to compute a residual. Figure 10a shows the correlation between the source position and the position measured using the pads. The difference between the two coordinates exhibits a differential non-linearity as shown in Fig. 10b. The periodicity is roughly half the pad size (~ 1 mm) due to the symmetry of the zigzag strips within the pad.

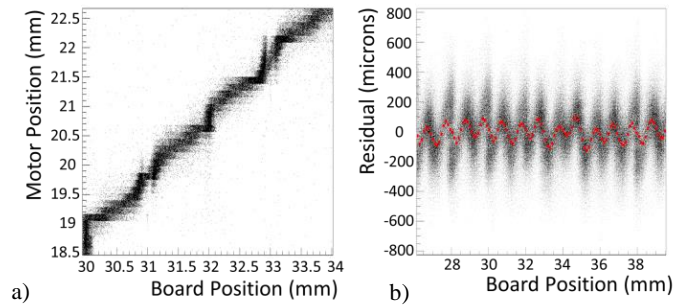

Fig. 10 a) Correlation between the position determined from the chevron pads and the actual source position measured across the pad plane with a high precision X-ray source b) Differential non-linearity of the chevron pads as a function of position across the pad plane.

Since the periodic structure for the pads is an intrinsic property of the readout board, the residuals can be used to correct the position determined from the centroid measurement for the differential non-linearity. This effect was measured in the lab and used to correct the test beam data. Figure 11a shows the correction function derived using this procedure for events with 2 pads and 3 pads hit. The correction is ~ ± 100 μm and covers a range of half the pad size, but was applied to the entire pad using symmetry. Figure 11b shows an example of the residual distribution for the position coordinate from the pads before and after applying this correction, which results in narrowing the distribution from 140 μm to 110 μm.

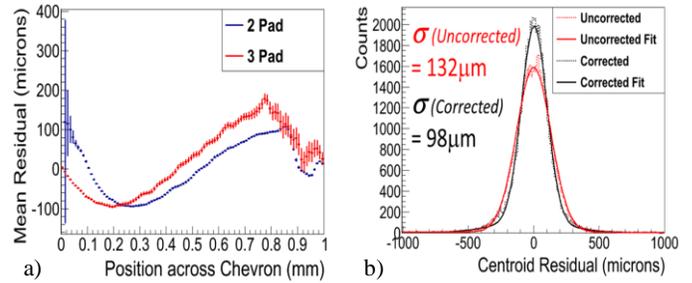

Fig. 11. a) Correction function for the position determined from the pads as a function of position within the pad (shown separately for events with 2 pads and 3 pads hit). b) Residual distribution before and after applying the differential non-linearity correction.

Figure 12 shows the position resolution of the minidrift detector with the chevron pads as a function of incident track angle. The light orange curve shows the resolution before correcting for the differential non-linearity of the chevrons and the dark orange curve shows the resolution after the correction. It starts at a value less than 100 μm at zero degrees after correction and reaches a value of slightly more than 400 μm at 45°. The differential non-linearity correction has a greater effect for angles less than 10° due to the fact that at these angles, one is relying on only a few pads to determine the centroid, and the interpolation within the pad plays a greater role in determining the overall position. The figure also shows the resolution after unfolding the contribution from the timing jitter. After this unfolding, the resolution decreases to ~ 350 μm at 45°. Also shown for comparison is the resolution for the COMPASS strips for the average of the X and Y planes for the orthogonal position, both before and after unfolding the contribution from the timing jitter. As expected, the strips provide better resolution at all angles, but the pads still provide a resolution of < 150 μm for angles < 30°, yet with a much coarser segmentation that can be used with higher hit densities.

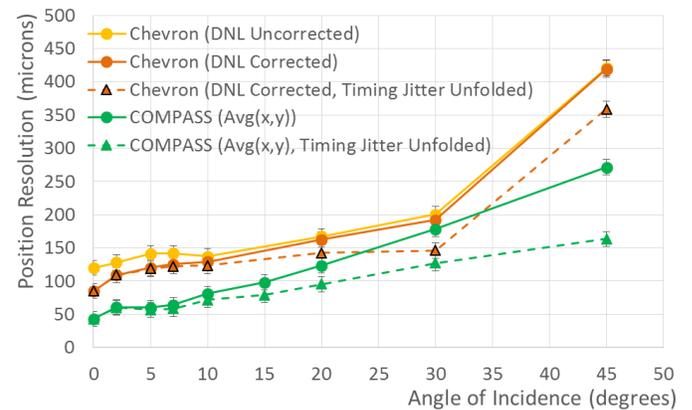

Fig. 12 Position resolution for the chevron pad readout as a function of angle. Curves are shown before and after correcting for the differential non-linearity of the chevrons and after unfolding the contribution from the timing jitter from the corrected curve. Also shown is the resolution for the average of the XY strips from the COMPASS readout before and after unfolding the contribution from the timing jitter.

Figure 13 shows the angular resolution for the chevron pad readout. The resolution is ~ 30 mrad at zero degrees and decreases to < 20 mrad for angles greater than 5°. Also shown is the angular resolution of COMPASS strips (average of the X and Y planes in the orthogonal orientation) for comparison. The angular resolution of the chevrons is considerably worse than the strips at smaller angles due to the fact that the strips provide more discrete points with better resolution for fitting the vector at smaller angles. For less than 5°, only one or two chevrons are hit, which limits the ability to determine the angle with high precision. However, for angles greater than 30° the resolution of the chevrons and strips are quite comparable.

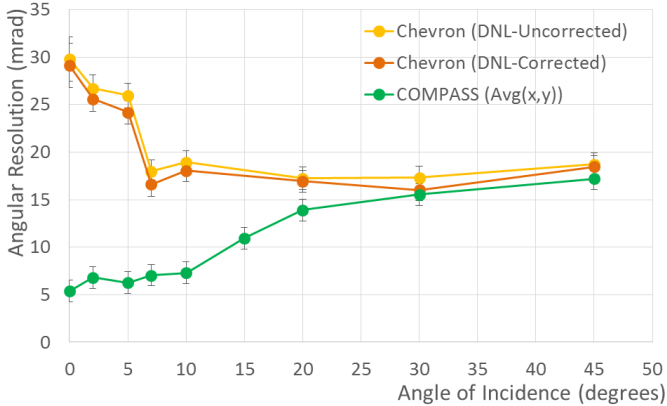

Fig. 13 Angular resolution for the chevron pad readout as a function of angle. The light orange curve is uncorrected and the dark orange curve is corrected for the differential non-linearity. Also shown is the angular resolution for the COMPASS readout strips for the X view.

*4.3 Comparison with the rising edge method*

All of the results shown above were obtained with the time slice method for vector reconstruction. However, the rising edge method, where the arrival time of the charge on the electrode is determined from the rising edge of the preamp pulse, is often used for this type of TPC detector. We therefore give a comparison of the results obtained with the rising edge method (also described in Refs. [6,7]) with those obtained with the time slice method.

Figure 14 shows a comparison of the position resolution with the two methods. The red curve gives the results using the rising edge method to measure a vector at all angles. Due to the fact that timing is not well determined (essentially to within one 25 ns time bin), and only a few strips can be used to calculate the vector, it gives very poor results at small angles. It then approaches the same value as for the time slice method for angles greater than 10°. However, at small angles, one can use a hybrid method for which a simple charge weighted centroid is calculated for small angles, and combine it with the rising edge vector method at larger angles. We define two weighting coefficients, $w_c=(N_{cut}/N_{strip})^2$ and $w_v=(N_{strip}/N_{cut})^2$, where $N_{cut}$ is a constant related to the number of strips that fire. For events with a small number of strips firing, $w_c$ gives a higher weight to the centroid measurement, whereas for a larger number of strips firing, $w_v$ gives a higher weight to the vector measurement. We then define an overall weighting factor $w_{comb}= (w_c \cdot x_c + w_v \cdot x_v)/(w_c+w_v)$ which is used to compute the combined resolution. The combined resolution was optimized by varying the parameter $N_{cut}$ and the result is shown by the blue curve in the Fig. 14. The resolution starts out agreeing with the time slice method (shown in light green) at very small angles, then increases for angles between 2.5° and 10°, and then agrees with the time slice method again for larger angles. While the rising edge method can also give acceptable results over part of the angular range, the time slice method can give better results over the full angular range.

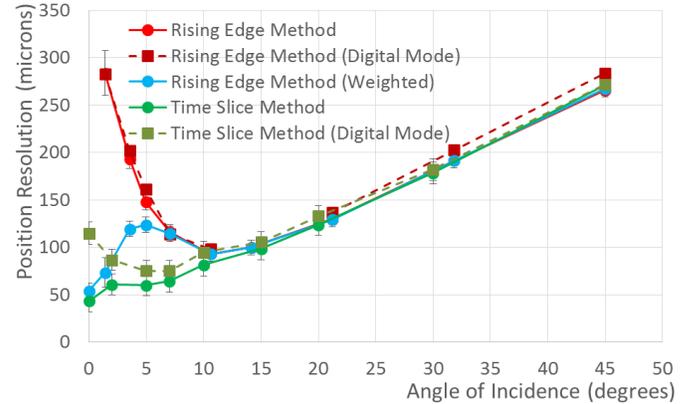

Fig. 14 Comparison of the position resolution for the COMPASS strips using the rising edge method versus the time slice method. The red curve is for the time slice method over the full angular range and the blue curve is for a weighted average of a simple centroid at small angles and the rising edge method at larger angles.

Also shown in Fig. 14 are results using a so-called "digital method". This method simply uses the center coordinates of the strips for the position (i.e., no centroid finding) and the average value of the time bin for either the rising edge of the pulse or the time bin for the time slice method. It is shown for illustrative purposes and shows that for both the rising edge method and the time slice method, there is essentially no improvement in the resolution for angles greater than 10° over using simply the digital information. The resolution in this case is dominated by the time resolution of one time bin. However, for the time slice method, there is a considerable advantage at smaller angles using the centroid information.

Figure 15 shows a comparison of the angular resolution of the rising edge and time slice methods. As mentioned above, the rising edge method gives a very poor determination of the vector for small angles where a small number of strips are hit. The angular resolution improves with increasing angle and reaches a value which is comparable to the time slice method for angles greater than 15°.

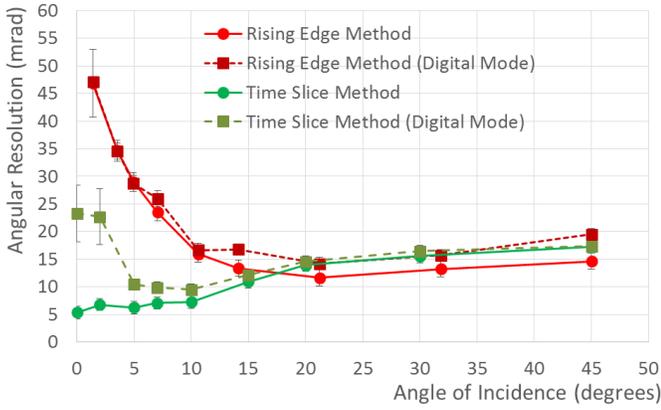

Fig. 15 Comparison of the angular resolution for COMPASS strips using the time slice method versus the rising edge method. The curves are for the average of the X and Y planes for the detector in the diamond position.

*4.4 Pulse shape deconvolution method*

It should be noted that the SRS readout system was not designed for TPC like applications, and the time resolution and shaping times for this system inherently limit the position and angular resolutions than can be achieved. However, one can attempt to overcome this limitation by using the knowledge of the pulse shape and time structure to unfold these effects from the final results. An alternative method was studied that attempted to reconstruct the original charge deposition pattern on each electrode in each time slice from the measured pulse shape using the known impulse response of the APV25 preamplifier. The technique, based on CPU-intensive non-linear minimization algorithms, accounts for the measured signal correlations in the neighboring time slices in a controlled way, and therefore reduces the confounding effects of the APV25 long shaping time. An example of this unfolding procedure applied to an inclined track in the detector is shown in Fig. 16 for the COMPASS style readout. It is clear that the effects of the long tails of the preamp pulse which affect neighboring strips in each time slice have been significantly reduced.

The subsequent clustering algorithm in each time slice not only uses more realistic values for the original charge deposition on each strip, but it also uses the full covariance matrix after the unfolding procedure to provide information on the correlations between each measurement. This technique potentially allows one to significantly improve the spatial resolution of the detector, especially at large incident track angles.

One can use the output of this algorithm to illustrate the effect of the SRS system timing jitter on the measured data. It is easy to demonstrate that this jitter should affect the GEM coordinate measurements in the X and Y planes in a correlated way for any given event. To be specific, in the diamond configuration, with each of the X and Y planes inclined by an angle $\alpha$ with respect to the track direction, a random timing offset $\Delta t$ between the trigger and the SRS internal 40 MHz clock should result in a systematic shift in both measured coordinates by a value $v*\Delta t*\tan(\alpha)$, where $v$ is the drift velocity of the electron charge cloud. The apparent residuals between the coordinates measured by the GEM detector and the reference silicon tracker should therefore be strongly correlated in the X and Y projections. This correlation is clearly visible in Fig. 17a for the COMPASS readout in the diamond configuration at 45°. It can be seen that while the residual width in both the X and Y planes separately is large and is dominated by the SRS timing jitter, the projection along the diagonal, shown in Fig 17b, which removes the correlated errors between the two measurements, shows a resolution that is well below 100 μm. However, it should be noted that this procedure eliminates all correlated errors between the two measurements, including fluctuations in the gas ionization process, etc., and therefore the projected resolution represents a lower limit to what resolution could actually be achieved in a real detector. However, we believe that the timing jitter of the SRS system is the dominant source of the correlated error in these measurements, and that a position resolution on the order of 100 μm should be achievable even at large angles. Further studies on the use of this algorithm will continue and will be described in more detail in a future publication.

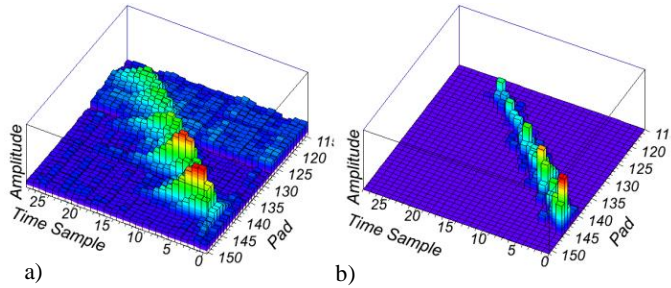

Fig. 16 a) Raw data for a single event (COMPASS readout, X plane in the "diamond" configuration at 45°. b) Same event after pedestal subtraction and charge distribution unfolding

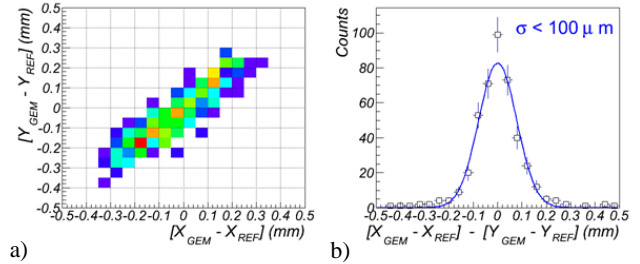

Fig. 17 a) 2D correlation between the GEM residuals with respect to the reference silicon telescope in the X and Y projections for the COMPASS readout in the diamond configuration at 45°. b) Projected resolution along the diagonal in Fig 17a where correlations between the X and Y coordinate measurements cancel out.

## V. DISCUSSION

The main purpose of this study was to investigate how measuring a vector for a track with a GEM detector would improve its position resolution at larger angles, and to measure the angular resolution that could be obtained from measuring the vector in a single detector. This was motivated by the fact that in many applications, the tracking system needs to measure a range of angles from normal incidence up to approximately 45 degrees, and the position resolution for a normal planar GEM tracking detector deteriorates rapidly at larger angles. Also, tracking systems typically have multiple detectors separated by some distance in order to measure the angle or bending angle of the track. However, each detector adds material in the particle's path, which increases the multiple scattering and the probability for secondary interactions. In the case of electrons, it also increases the probability for bremsstrahlung radiation, which worsens the momentum and energy resolution. By improving the resolution at each measuring station and also measuring the track angle, it is possible to reduce the number of detectors in the tracking system and still obtain good position and momentum resolution.

Figure 8 shows that one can obtain a position resolution of less than 300 μm up to an angle of 45° using the vector information from the GEM. This could be reduced to ~ 150 μm with a readout system that provides better time resolution, and perhaps to ~ 100 μm with proper treatment of the tails of the preamp pulse. This can improve the performance of a large or wide angle tracking system with fewer detectors with less material and at a lower cost. Figure 9 shows that the angular resolution is less than 20 mrad up to 45°, which may not be adequate for a precise determination of the particle's momentum in a final analysis, but may be sufficient for triggering purposes. One such application using minidrift micromegas detectors is described in Ref. [10].

Figures 12 and 13 show that excellent position and angular resolution can be obtained with the chevron pads. The chevrons can offer significant advantages in many applications where one needs to minimize the channel count or where a high degree of segmentation is required due to high particle multiplicity. It can also be used as a TPC readout where a large number of individual pads are required that only require a high precision coordinate determination in one direction (e.g., r-φ), such as described in Refs. [4,5].

The measurements carried out in this study were performed with a rather limited set of parameters which could be further optimized to improve the detector's performance. These include varying the size of the drift gap, using a different operating gas which would increase the diffusion and charge spread over the pads, varying the size and shape of the chevron pads, and improving the readout electronics. In particular, the SRS system was not specifically designed for this type of application, and electronics with better time resolution and a larger dynamic range would provide a much better measure of the drift time and total charge. In addition, improvements can be made in the algorithms used for determining the vector that could achieve a position resolution ~ 100 μm at large angles.

Finally, it should be noted that virtually any planar GEM detector can be converted to a minidrift type detector by simply increasing the drift gap. The only additional requirement would be to implement the ability to measure the drift time in the readout electronics. This would allow improving the resolution of any detector that requires measuring tracks at larger angles. At the EIC, the spectrometer in the hadron going direction will have a number of detectors that will need to measure tracks at large angles, including large angle GEM trackers and particle ID detectors. These devices are currently being studied by other groups that are part of a consortium that is developing tracking detectors for EIC, and this work has been a part of that effort. Further details and additional results from these studies can be found in Refs [11,12,13].

## VI. CONCLUSIONS

This study investigated the properties of a GEM detector with a small drift region that was operated in a TPC like mode that allowed the determination of a track vector in a single detector. With a drift gap ~ 1-2 cm, it was shown that it is possible to reconstruct a vector which can be used to improve the position resolution for larger incident track angles, and also provide moderate angular resolution. Two different readout methods were studied, one using narrow strips and another using chevron pads. It was shown that the chevron pads can provide good resolution for both position and angle, and can be used in a high multiplicity environment. This type of minidrift GEM detector is being developed for future use at the Electron Ion Collider as well as other applications for tracking at large angles.

## VII. ACKNOWLEDGMENT

We would like to thank L. Uplegger, R. Rivera and L. Vigani for their help and support in operating the silicon telescope in the Fermilab MT6 test beam area and the analysis of its data that enabled us to measure the incoming beam track with very high precision.

## VIII. REFERENCES

[1] C. Altunbas et.al., "Construction, test and commissioning of the triple Gem tracking detector Compass", Nucl. Instr. Meth. Vol. A490 (2002) 177-203.
[2] T.Miki, R.Itoh and T.Kamae, "Zigzag-shaped pads for cathode readout of a time projection chamber", Nucl. Instr. Meth. Vol A236 (1985) 64-68.
[3] E.Mathieson and G.C.Smith, "Reduction in non-linearity position sensitive MWPCs", IEEE Trans. Nucl. Sci. Vol. 36. No. 1 (1989) 305-310.
[4] B.Yu et.al., "Study of GEM Characteristics for Application in a MicroTPC", IEEE Trans. Nucl. Sci. Vol. 50, No. 4 (2003) 836-841.
[5] B.Yu et.al.,"A GEM Based TPC for the LEGS experiment", Conference Record Proceedings of the 2005 IEEE Nuclear Science Symposium and Medical Imaging Conference, October 23-29, 2005, Fajardo, Puerto Rico.
[6] M.L. Purschke, "Test Beam Study of a Short Drift GEM Tracking Detector" Conference Record Proceedings of the 2013 IEEE Nuclear Science Symposium and Medical Imaging Conference, October 27-Nov 2, 2012, Seoul, Korea.

[7]  T.Cao et.al., "A Study of a GEM Tracking Detector for Imaging Positrons from PET Radioisotopes in Plants", IEEE Trans. Nucl. Sci. Vol. 61, No. 5 (2014) 2464-2471.
[8]  Scalable Readout System (SRS) developed by the CERN RD51 Collaboration, M. Chefdeville et.al., "RD51: A World Wide Collaboration for the Development of Micropattern Gas Detectors", J. Phys. Conf. Ser. 309:012017 (2011).
[9]  M.L. Purschke, "RCDAQ: A New Readout System for the CERN SRS Readout Electronics", Conference Record Proceedings of the 2012 IEEE Nuclear Science Symposium and Medical Imaging Conference, October 29-Nov 3, 2012, Anaheim, CA.
[10] Y.Kataok, S.Leontsinis and K.Ntekas, "Performance studies of a micromegas chamber in the ATLAS environment", Proceedings of the 3$^{rd}$ International Conference on Micropattern Gas Detectors, JINST 9 (2014) C03016.
[11] Generic Detector R&D for an Electron Ion Collider, eRD6 EIC Tracking Consortium, EIC Wiki https://wiki.bnl.gov/conferences/index.php/EIC-Detector-Proposals
[12] M.Blatnik et.al., "Performance of a Quintuple GEM Based RICH Detector Prototype", arXiv:1501.03530 (submitted to IEEE Trans. Nucl. Sci.)
[13] K.Gnanvo et.al., "Large Size GEM for Super Bigbite Spectrometer (SBS) polarimeter for Hall A 12 GeV program at JLAB", Nucl. Instr. Meth. A782 (2015) 77-86.